\documentclass[floatfix, preprint, showpacs, showkeys, preprintnumbers, nofootinbib, superscriptaddress]{revtex4-1}
\usepackage[utf8]{inputenc}
\usepackage[sort&compress]{natbib}
\usepackage{ulem}
\usepackage{bm}
\usepackage{times}
\usepackage{amssymb,amsbsy,amsmath,amsfonts}
\usepackage{graphicx}
\usepackage{epstopdf}
\usepackage{float}
\usepackage{color}
\usepackage{morefloats}
\usepackage{rotating}
\usepackage{srcltx}
\usepackage{slashed}
\usepackage{subfigure}
\usepackage{multirow}
\usepackage{verbatim}
\usepackage{hyperref}
\usepackage{overpic}
\usepackage{tabularx}

\begin{document}

\title{ Masses and strong decays  of open charm hexaquark states $\Sigma_{c}^{(\ast)}{\Sigma}_{c}^{(\ast)}$}

\author{Xi-Zhe Ling}
\affiliation{School of Physics, Beihang University, Beijing 102206, China}

\author{Ming-Zhu Liu}\email{zhengmz11@buaa.edu.cn}
\affiliation{School of Space and Environment, Beihang University, Beijing 102206, China}
\affiliation{School of Physics, Beihang University, Beijing 102206, China}

\author{Li-Sheng Geng}\email{lisheng.geng@buaa.edu.cn}
\affiliation{School of Physics, Beihang University, Beijing 102206, China}
\affiliation{
Beijing Key Laboratory of Advanced Nuclear Materials and Physics,
Beihang University, Beijing 102206, China}
\affiliation{School of Physics and Microelectronics, Zhengzhou University, Zhengzhou, Henan 450001, China}

\date{\today}
\begin{abstract}
Inspired by the recent discovery of the doubly charmed tetraquark state $T_{cc}^{+}$ by the LHCb Collaboration,
we perform a systematic study of  masses and strong decays of open charm hexaquark states ${\Sigma}_{c}^{(\ast)}\Sigma_{c}^{(\ast)}$. Taking into account heavy quark spin symmetry breaking,  we predict several bound states of  isospin $I=0$, $I=1$, and $I=2$ in the one boson exchange model.  Moreover, we  adopt the effective Lagrangian approach to estimate the decay widths of ${\Sigma}_{c}^{(\ast)}\Sigma_{c}^{(\ast)} \to \Lambda_{c}\Lambda_{c}$ and their relevant ratios via the triangle diagram mechanism, which range from a few MeV to a few tens of MeV. We strongly recommend future experimental searches for the ${\Sigma}_{c}^{(\ast)}\Sigma_{c}^{(\ast)}$ hexaquark states in the $\Lambda_c\Lambda_c$ invariant mass distributions.

\end{abstract}


\maketitle

\section{Introduction}

The quenched (or conventional) quark model can  well describe the properties of traditional hadrons, especially  the ground-state ones~\cite{Godfrey:1985xj,Capstick:1985xss}. However, a large number of  states that can not be easily explained by the quenched quark model have been accumulating  since  $D_{s0}^*(2317)$  and $X(3872)$ were discovered by the BaBar ~\cite{BaBar:2003oey} and  Belle~\cite{Belle:2003nnu}  collaborations in 2003. To unveil the nature of these exotic states, many pictures, such as multiquark states, hadronic molecules, and kinetic effects, have been proposed  to understand them from different perspectives,  including production mechanisms, mass spectra, and decay widths~\cite{Chen:2016qju,Hosaka:2016ypm,Lebed:2016hpi,Guo:2017jvc,Olsen:2017bmm,Ali:2017jda,Brambilla:2019esw,Liu:2019zoy}.
Among them,  the hadronic molecular picture is probably the most popular one since most of these exotic states are located close to the mass thresholds of  conventional hadron pairs.   Two-body hadronic molecules are expected to exist in three configurations, i.e., meson-meson, meson-baryon, and baryon-baryon. With more and more $XYZ$ states, $P_{c}$ states, and $T_{cc}$ discovered  in the past two  decades~\cite{Brambilla:2019esw},  heavy hadronic  molecules have attracted a lot attention.

Heavy quark spin symmetry (HQSS) plays an important role in describing heavy hadronic molecules~\cite{Nieves:2012tt,Guo:2013sya,Xiao:2013yca,Liu:2019tjn}. HQSS dictates that the interaction between a light (anti)quark and a heavy (anti)quark  is independent of the heavy (anti)quark spin in the limit of heavy quark masses.  As the heavy mesons ($D$, $D^{\ast}$) and baryons ($\Sigma_{c}$, $\Sigma_{c}^{\ast}$) belong to the same HQSS doublets,  using them as building blocks, one expects three  hidden charm multiplets of hadronic molecules, i.e.,  $\bar{D}^{(\ast)}D^{(\ast)}$, $\bar{D}^{(\ast)}\Sigma_{c}^{(\ast)}$, and $\bar{\Sigma}_{c}^{(\ast)}\Sigma_{c}^{(\ast)}$. With the assumption  that $X(3872)$ is a $\bar{D}^{\ast}D$  bound state, in Ref.~\cite{Guo:2013sya}, Guo et al.  adopted an effective field theory approach to study the $\bar{D}^{(\ast)}D^{(\ast)}$ system and predicted a $\bar{D}^{\ast}D^{\ast}$ bound state as the HQSS partner of $X(3872)$.  In Ref.~\cite{Liu:2019tjn}, using the same approach we obtained a complete HQSS multiplet of hadronic molecules, $\bar{D}^{(\ast)}\Sigma_{c}^{(\ast)}$, where  $P_{c}(4440)$ and $P_{c}(4457)$ are regarded as the $\bar{D}^{\ast}\Sigma_{c}$ bound states. Although    the likely existence of   $\bar{\Sigma}_{c}^{(\ast)}\Sigma_{c}^{(\ast)}$ bound states have been explored theoretically~\cite{Li:2012bt,Lu:2017dvm,Peng:2020xrf,Dong:2021juy}, they have not been observed experimentally.

Very recently, the LHCb Collaboration observed one open charm tetraquark state $T_{cc}^{+}$  below the $D^{0}D^{\ast+}$ mass threshold~\cite{LHCb:2021vvq}, which has already been  predicted by a lot of theoretical works in the diquark and anti-diquark  picture~\cite{Ader:1981db,Zouzou:1986qh,Lipkin:1986dw,Heller:1986bt,Carlson:1987hh,Silvestre-Brac:1993zem,Semay:1994ht,Gelman:2002wf,Vijande:2003ki,Yang:2009zzp,Luo:2017eub,Karliner:2017qjm,Mehen:2017nrh,Eichten:2017ffp} as well as in the molecular picture~\cite{Manohar:1992nd,Pepin:1996id,Janc:2004qn,Molina:2010tx,Li:2012ss,Feng:2013kea,Wang:2017uld,Junnarkar:2018twb,Maiani:2019lpu,Liu:2019stu}. In our previous work~\cite{Liu:2019stu}, using the one boson exchange (OBE) model we have systematically studied the ${D}^{(\ast)}D^{(\ast)}$ system and predicted one isocalar $DD^{\ast}$ molecule with a binding energy of about 3 MeV, which is an open charm tetraquark state and may correspond to the $T_{cc}^{+}$ state.
Therefore, it is interesting to investigate the likely existence of open charm pentaquark and hexaquark states. In Ref.~\cite{Liu:2020nil}, we  obtained a complete HQSS multiplet of hadronic molecules ${D}^{(\ast)}\Sigma_{c}^{(\ast)}$, consistent with Refs.~\cite{Chen:2021htr,Dong:2021bvy,Chen:2021kad}. One should note that the minimum quark content of the ${D}^{(\ast)}\Sigma_{c}^{(\ast)}$ system is $ccq$, which  may  couple to the excited states of $\Xi_{cc}$.
Consequently, the  ${D}^{(\ast)}\Sigma_{c}^{(\ast)}$ molecules can mix with the  excited  doubly charmed baryons. In the present work,  we  investigate the mass spectrum and strong decays of  ${\Sigma}_{c}^{(\ast)}\Sigma_{c}^{(\ast)}$, which cannot couple to conventional baryons of $ccq$.

Up to now, the deuteron is the only experimentally confirmed baryon-baryon bound state, whose property can  be well reproduced by high precision nuclear forces~\cite{Machleidt:1987hj,Machleidt:2000ge}.
It is natural to expect the existence of other dibaryon  states  consisting of strange or charm quarks.  In 1976, Jaffe predicted one famous dibaryon state $\Lambda\Lambda$~\cite{Jaffe:1976yi}, who has received a lot of attention  both experimentally and theoretically~\cite{Balachandran:1983dj,Takahashi:2001nm,Polinder:2007mp,Yoon:2007aq,Inoue:2010es,NPLQCD:2010ocs,Morita:2014kza,Li:2018tbt}.   However, the existence of the $\Lambda\Lambda$ dibaryon is still controversial.
In recent years,  the  theoretical and experimental studies of dibaryon states have been extended to systems containing more strangeness quarks~\cite{Haidenbauer:2014rna,Morita:2016auo,Gongyo:2017fjb,Sasaki:2017ysy,Li:2018tbt,Sekihara:2018tsb,STAR:2018uho,Huang:2019esu,Chen:2019vdh,Morita:2019rph,Hohlweger:2020cpl,Liu:2020uxi,Ogata:2021mbo}.

Experimental searches for dibaryons containing charm quarks is  challenging because of the high energy required and the low production of charmed baryons. On the other hand, many theoretical studies have been performed.   In Ref.~\cite{Liu:2011xc}, Liu et al.  predicted one bound state generated by $\Lambda_{c}N$ and $\Sigma_{c}N$ couple channels in the OBE model, which was later explored in the chiral constituent quark model~\cite{Huang:2013zva,Gal:2014jza},   lattice QCD simulations~\cite{Miyamoto:2017tjs}, and  effective field theories~\cite{Song:2020isu}.  Later, using the same approach, Li et al. systematically studied possible bound states made of two singly charmed baryons~\cite{Lee:2011rka,Li:2012bt,Yang:2018amd}. Some of these states  were investigated in the quark delocalization color screening model~\cite{Huang:2013rla,Xia:2021nif}.    After the discovery of the doubly charmed baryon $\Xi_{cc}$ by the LHCb Collaboration~\cite{LHCb:2017iph}, Meng et al.  investigated bound states made of two doubly charmed baryons~\cite{Meng:2017fwb}. Inspired by  the discovery of the pentaquark states by the LHCb Collaboration~\cite{LHCb:2019kea}, we  studied the $\Xi_{cc}^{(\ast)}\Sigma_{c}^{(\ast)}$ system utilizing heavy antiquark diquark symmetry in the OBE model and effective field theory~\cite{Pan:2019skd,Pan:2020xek} and obtained a complete HQSS multiplet of hadronic molecules.   Recently, lattice QCD simulations obtained a series of dibaryon states with open charm, such as  $\Xi_{cc}\Sigma_{c}$~\cite{Junnarkar:2019equ} and  $\Omega_{ccc}\Omega_{ccc}$~\cite{Lyu:2021qsh}.

In this work,  we investigate the mass spectrum  and decay patterns of the  $\Sigma_{c}^{(\ast)}\Sigma_{c}^{(\ast)}$ system dictated by HQSS. In Ref.~\cite{Peng:2020xrf}, we parameterized  the contact-range potential of the  $\Sigma_{c}^{(\ast)}\Sigma_{c}^{(\ast)}$ system with three parameters. However, we can not fully determine the mass spectrum of the $\Sigma_{c}^{(\ast)}\Sigma_{c}^{(\ast)}$ system because no experimental data is available. In this work, we turn to  the OBE model for help and  calculate the mass spectrum.  Then based on the so-obtained mass spectrum, we estimate the strong decay widths of   $\Sigma_{c}^{(\ast)}\Sigma_{c}^{(\ast)}\to \Lambda_{c}\Lambda_{c}$ through the effective Lagrangian approach, which has been widely used to explore the strong decays of hadronic molecules~\cite{Faessler:2007gv,Dong:2008gb,Xiao:2019mvs,Dong:2019ofp}.

  The manuscript is structured as follows. In Sec.~\ref{sec:obe} we present the details of the OBE
model as applied to the  $\Sigma_{c}^{(\ast)}\Sigma_{c}^{(\ast)}$ system and the numerical results of their mass spectra.
In Sec.~\ref{sec:pre} we explain the effective Lagrangian approach and calculate the decay widths of $\Sigma_{c}^{(\ast)}\Sigma_{c}^{(\ast)} \to\Lambda_{c}\Lambda_{c}$.
Finally we conclude in Sec.~\ref{sum}

\section{Mass spectrum of $\Sigma_{c}^{(\ast)}\Sigma_{c}^{(\ast)}$  system  }
\label{sec:obe}

\subsection{OBE potential}
In this work,  we employ the OBE model to derive the  $\Sigma_{c}^{(\ast)}\Sigma_{c}^{(\ast)}$ interaction, which has been successfully  used to investigate  hadronic molecules composed of heavy hadrons~\cite{Swanson:2003tb,Ding:2008gr,Liu:2008fh,Yang:2011wz}.
The OBE interaction
of two heavy hadrons is generated by the exchange of light mesons,  $\pi$, $\sigma$, $\rho$, and $\omega$. The vector
mesons, $\rho$ and $\omega$, provide the short-range interaction, the scalar meson $\sigma$ provides the medium-range
interaction, and the $\pi$ meson provides the long-range interaction. Given the exploratory nature of
the present work, the contributions of other mesons are neglected~\cite{Liu:2019zvb}.
To derive the OBE potential, the effective Lagrangians describing the interaction between a charmed baryon and a light meson are necessary.   The relevant
Lagrangians are the same as those of the
$\bar{D}^{(\ast)}\Sigma_{c}^{(\ast)}$ interaction~\cite{Liu:2019zvb}, which read 
\begin{eqnarray}
    \mathcal{L}_{S_{c}S_{c}\pi} &=& \frac{ig}{\sqrt{2} f_{\pi}}
{\vec{S}_{c}}^{\dagger} \cdot ( \vec{T} \cdot \vec{\pi}\vec{\nabla}
 \times \vec{S}_{c}) \, , \label{eq:L-pi} \\
  \mathcal{L}_{S_{c}S_{c}\sigma} &=& g_{\sigma}
    {\vec{S}_{c}}^{\dagger} \sigma\cdot  \vec{S}_{c} \, , \label{eq:L-sigma} \\
  \mathcal{L}_{S_{c}S_{c} \rho} &=& g_{\rho}
    {\vec{S}_{c}}^{\dagger} (\vec{T} \cdot \vec{\rho}^{0}) \cdot \vec{S}_{c}
  \nonumber \\
  &-& \frac{f_{\rho }}{4 M}\,
    {\vec{S}_{ci}}^{\dagger}  \vec{T} \cdot \left( \partial^i \vec{\rho}^j
    - \partial^j \vec{\rho}^i \right)\vec{S}_{cj}  \, , \label{eq:L-rho} \\
    \mathcal{L}_{S_{c}S_{c}\omega} &=& g_{\omega}
    {\vec{S}_{c}}^{\dagger} {\omega}^{0}\cdot \vec{S}_{c}
  \nonumber \\
  &-& \frac{f_{\omega}}{4 M}\,
    {\vec{S}_{ci}}^{\dagger} \, \left( \partial^i {\omega}^j
    - \partial^j {\omega}^i \right) \vec{S}_{cj}  \, , \label{eq:L-omega}
\end{eqnarray}
where $\vec{S}_{c}=(\frac{1}{\sqrt{3}}\Sigma_{c}\vec{\sigma}+\vec{\Sigma}_{c}^{\ast} )$ denotes the superfield of $\Sigma_{c}$ and $\Sigma_{c}^{\ast}$ dictated by HQSS. The coupling of the $\pi$ meson to $\Sigma_{c}$ and $\Sigma_{c}^{\ast}$,
$g$, was determined to be $0.84$ in lattice QCD~\cite{Detmold:2012ge}, which is smaller than the prediction of the quark model~\cite{Liu:2011xc}. In this work we take $g=0.84$.
For the coupling to the $\sigma$ meson, we estimate it using the
 quark model.  From the nucleon and $\sigma$ meson coupling of the
 liner sigma model, $g_{\sigma NN}=10.2$, the corresponding coupling  is determined to  be  $g_{\sigma}=\frac{2}{3}g_{\sigma NN}=6.8$~\cite{GellMann:1960np}.
The couplings of the light vector mesons contain both
electric-type($g_{v}$) and magnetic-type($f_{v}$) ones, which are related via $f_{v}=\kappa_{v}g_{v}$.
For the $\rho$ and $\omega$ couplings, from SU(3)-symmetry and the OZI rule, we obtain $g_{\omega}=g_{\rho}$ and $f_{\omega}=f_{\rho}$.
The electric-type coupling of the singly charmed baryons to the vector mesons
is determined to be $g_{v}=5.8$~\cite{Liu:2011xc} as well as the corresponding magnetic-type coupling  is estimated to be $\kappa_{v}=1.7$~\cite{Can:2013tna}. All the couplings are given in Table \ref{tab:couplings2} for easy reference.

\begin{table}[ttt]
\centering
\caption{Couplings of charmed baryons to light mesons. The magnetic coupling of the $\rho$ and $\omega$ mesons are defined as $f_{v}=\kappa_{v}g_{v}$, and
$M$ refers to the mass scale (in units of MeV) involved in the magnetic-type couplings.
}
\label{tab:couplings2}
\begin{tabular}{c|c c c c c c c}
  \hline \hline
     Coupling  &~~~~  $g$   &~~~~  $g_{\sigma }$ &~~~~ $g_{\rho }$ &~~~~ $g_{\omega }$  &~~~~ $\kappa_{\rho }$ &~~~~  $\kappa_{\omega }$ &~~~~  $M$
     \\ \hline  Value   &~~~~  0.84 &~~~~ 6.8 &~~~~ 5.8 &~~~~ 5.8 &~~~~ 1.7&~~~~ 1.7 &~~~~ 940  \\
  \hline \hline
\end{tabular}
\end{table}

 With the above Lagrangians, the OBE potentials for the $\Sigma_{c}^{(*)}\Sigma_c^{(*)}$ system in coordinate space can be obtained,  which read
\begin{eqnarray}
  V_{\pi}(\vec{r}) &=&
  \vec{T}_{1} \cdot \vec{T}_{2}\,\frac{g^{2}}{6 f_{\pi}^2}\,\Big[
    - \vec{a}_{1} \cdot \vec{a}_{2}\,\delta(\vec{r})
    \nonumber \\ && \quad
    + \, \vec{a}_{1} \cdot \vec{a}_{2}\,m_{\pi}^3\,W_Y(m_{\pi} r)
    \nonumber \\ && \quad
    + \, S_{12}(\vec{r})\,m_{\pi}^3\,W_T(m_{\pi} r) \Big] \, , \label{11}  \\
  V_{\sigma}(\vec{r}) &=& -{g_{\sigma }^2}\,m_{\sigma}\,W_Y(m_{\sigma} r)
  \, ,  \label{22} \\
  V_{\rho}(\vec{r}) &=&\vec{T}_{1} \cdot \vec{T}_{2}\,\Big[
    {g_{\rho}^2}\,m_{\rho}\,W_Y(m_{\rho} r) \nonumber \\
    && \quad + \frac{f_{\rho}^2}{4 M^2}\,\Big(
    -\frac{2}{3}\,\vec{a}_{1} \cdot \vec{a}_{2} \delta(\vec{r})
    \nonumber \\ && \quad
    +\frac{2}{3}\,\vec{a}_{1} \cdot \vec{a}_{2}
    \, m_{\rho}^3 \, W_Y(m_{\rho} r)
    \nonumber \\ && \quad
    -\frac{1}{3}\,S_{12}(\hat{r})\, m_{\rho}^3 \, W_T(m_{\rho} r) \,\,
    \Big) \, \Big] \, , \label{33}  \\
    V_{\omega}(\vec{r}) &=&
    {g_{\omega}^2}\,m_{\omega}\,W_Y(m_{\omega} r) \nonumber \\
    && + \frac{f_{\omega}^2}{4 M^2}\,\,\Big[
    -\frac{2}{3}\,\vec{a}_{1} \cdot \vec{a}_{2} \, \delta(\vec{r})
    \nonumber \\ &&
    +\frac{2}{3}\,\vec{a}_{1} \cdot \vec{a}_{2} \,
    m_{\omega}^3 \, W_Y(m_{\omega} r)
    \nonumber \\ &&
    -\frac{1}{3}\,S_{12}(\hat{r})\, m_{\omega}^3 \, W_T(m_{\omega} r) \,\,
    \Big] \, ,  \label{44}
\end{eqnarray}
where  $\vec{T}_{1} \cdot \vec{T}_{2}$ is the isospin factor for the $\Sigma_{c}^{(\ast)}{\Sigma}_{c}^{(\ast)}$ system, and $\vec{a}_{1} \cdot \vec{a}_{2}$ and $S_{12}(\hat{r})$ denote the spin-spin and tensor terms, respectively, and $\vec{a}$ denotes the spin operator of  the $\Sigma_{c}^{(\ast)}$ baryon.
The dimensionless functions $W_Y(x)$ and $W_T(x)$ are defined as
\begin{eqnarray}
  W_Y(x) &=& \frac{e^{-x}}{4\pi x} \, , \\
  W_T(x) &=& \left( 1 + \frac{3}{x} + \frac{3}{x^2} \right)
  \,\frac{e^{-x}}{4\pi x} \, .
\end{eqnarray}

Since the charmed baryons and light mesons involved in our study are not point-like
particles, we introduce a form factor in the interaction vertices. Here we  use a monopolar form factor (for
more details we refer to  Refs.~\cite{Liu:2019stu,Liu:2019zvb})
\begin{eqnarray}
  F(q, m_{E}, \Lambda)=\frac{\Lambda^{2}-m_{E}^2}{\Lambda^{2}-{q}^2} \label{Eq:FF} \,
  ,
\end{eqnarray}
where $m_{E}$ and $q$ are the mass and 4-momentum of the exchanged
meson.  $\Lambda$ is an unknown cutoff, which is often fixed by reproducing some  hadronic molecular candidates that can be related to hadronic molecules of interest via  symmetries. Assuming  $X(3872)$ and $P_{c}(4312)$ as  bound states of $\bar{D}D^{\ast}$ and $\bar{D}\Sigma_{c}$, we can determine the corresponding cutoff of the OBE model to be
1.04 GeV~\cite{Liu:2019stu} and 1.12 GeV~\cite{Liu:2019zvb}, respectively.
In addition, the cutoff needed to reproduce the binding energy of the deuteron is  0.86 GeV~\cite{Liu:2021pdu}. All these results indicate that a reasonable value for the cutoff of the OBE model is about 1 GeV. In this work, assuming that the  $\Sigma_{c}^{(\ast)}$-$\Sigma_{c}^{(\ast)}$ system is similar to the nucleon-nucleon system,  we  take  a  cutoff of $\Lambda=$0.86 GeV to calculate the binding energies of the ${\Sigma}_{c}^{(\ast)}\Sigma_{c}^{(\ast)}$ dibaryon system,

With the above form factor, the functions $\delta$, $W_{Y}$, and
$W_{T}$ in Eqs.~(\ref{11}-\ref{44}) need to be changed accordingly
\begin{eqnarray}
  \delta(r) &\to& m^3\,d(x,\lambda) \, , \\
  W_Y(x) &\to& W_Y(x, \lambda) \, , \\
  W_T(x) &\to& W_T(x, \lambda) \, ,
\end{eqnarray}
with $\lambda = \Lambda / m$.
The corresponding functions $d$, $W_Y$, and $W_T$ read
\begin{eqnarray}
  d(x, \lambda) &=& \frac{(\lambda^2 - 1)^2}{2 \lambda}\,
  \frac{e^{-\lambda x}}{4 \pi} \, , \\
  W_Y(x, \lambda) &=& W_Y(x) - \lambda W_Y(\lambda x) \nonumber \\ && -
  \frac{(\lambda^2 - 1)}{2 \lambda}\,\frac{e^{-\lambda x}}{4 \pi} \, , \\
  W_T(x, \lambda) &=& W_T(x) - \lambda^3 W_T(\lambda x) \nonumber \\ && -
  \frac{(\lambda^2 - 1)}{2 \lambda}\,\lambda^2\,
  \left(1 + \frac{1}{\lambda x} \right)\,\frac{e^{-\lambda x}}{4 \pi} \, .
\end{eqnarray}

The generic wave function of a baryon-baryon system reads
\begin{eqnarray}
  | \Psi \rangle = \Psi_{J M}(\vec{r}) | I M_I \rangle\, ,
\end{eqnarray}
where $| I M_I \rangle$ denotes the isospin wave function and   $\Psi_{J M}(\vec{r})$ are the spin and spatial wave functions.
The dynamics in isospin space is embodied in the isospin factor $\vec{T}_{1}\cdot\vec{T}_{2}$.  In this work the total isospin is either 0, 1,  or 2  for the  ${\Sigma}_{c}^{(\ast)}\Sigma_{c}^{(\ast)}$ system,  then the corresponding isospin factor are $-2$, $-1$, and 1, respectively.

The spin wave function can be written as a sum over partial wave functions, which reads (in
the spectroscopic notation)
\begin{eqnarray}
  |{}^{2S+1}L_{J}\rangle &=& \sum_{M_{S},M_L}
  \langle L M_L S M_S | J M \rangle \, | S M_S \rangle \, Y_{L M_{L}}(\hat{r})
  \, , \nonumber \\
\end{eqnarray}
where $\langle L M_L S M_S | J M \rangle$ are the Clebsch-Gordan coefficients,
$| S M_S \rangle$ the spin wavefunction, and $Y_{L M_L}(\hat{r})$
the spherical harmonics.

In the partial wave decomposition of
the OBE potential, we encounter both spin-spin and tensor components
\begin{eqnarray}
  C_{12} &=& \vec{a}_1 \cdot \vec{a}_2 \, , \\
  S_{12} &=& 3 \vec{a}_1 \cdot \hat{r}\,\vec{a}_2 \cdot \hat{r} -
  \vec{a}_1 \cdot \vec{a}_2 \, ,
\end{eqnarray}
  In the present study, we
  consider both $S$ and $D$ waves.  The relevant matrix elements are listed
  in  Table~\ref{tab:spin1123}.

\subsection{Numerical results and discussions}

 For a pair of identical fermions, the total spin and isospin of  ${\Sigma}_{c}^{(\ast)}\Sigma_{c}^{(\ast)}$  have to fulfill  the condition, $I+S$=even.  The possible combinations of spin and isospin are shown in Table~\ref{dk2}.  Additionally, when deriving the OBE potential, we have assumed HQSS. As is well known, the charm quark mass is not large enough to strictly satisfy HQSS. In the present work, we take into account a  HQSS breaking of the order of $\Delta=15\%$  in the following way~\cite{Pan:2019skd}:
\begin{eqnarray}
V=V_{OBE}(1\pm\Delta).
\end{eqnarray}

With the above OBE potentials, the binding energies of the ${\Sigma}_{c}^{(\ast)}\Sigma_{c}^{(\ast)}$ system can be obtained by solving the Schr\"{o}dinger equation.
With  a cutoff of $\Lambda=0.86$ GeV, we obtain the binding energies of the ${\Sigma}_{c}^{(\ast)}\Sigma_{c}^{(\ast)}$ system for isospin $I=0$, $1$,  and $2$, which are given in Table~\ref{dk2}. After taking into account the HQSS breaking,
 we find  four hadronic molecules with $I=0$, i.e., $J^{P}=0^{+}$ ${\Sigma}_{c}\Sigma_{c}$, $J^{P}=1^{+}$ ${\Sigma}_{c}^{\ast}\Sigma_{c}^{\ast}$, $J^{PC}=0^{+}$ ${\Sigma}_{c}^{\ast}\Sigma_{c}^{\ast}$, and $J^{PC}=2^{+}$ ${\Sigma}_{c}^{\ast}\Sigma_{c}^{\ast}$.  Among them, the $J^{P}=0^{+}$ ${\Sigma}_{c}\Sigma_{c}$ bound state  corresponds to the deuteron counterpart in the charm sector.
 For isospin $I=1$, we  obtain four hadronic molecules as well but with smaller binding energies than those of  $I=0$. On the other hand, there exists only one bound state $I(J^{P})=2(2^{+})\Sigma_{c}^{\ast}\Sigma_{c}$ for isospin $I=2$. Obviously, with increasing isospin, the isospin factor decreases, which reduces  the number and binding energies  of ${\Sigma}_{c}^{(\ast)}\Sigma_{c}^{(\ast)}$ hadronic molecules.

 \begin{table}[ttt]
\centering
\caption{Binding energies of ${\Sigma}_{c}^{(\ast)}\Sigma_{c}^{(\ast)}$ molecules with the uncertainties originating from the breaking of HQSS and the corresponding masses, for which only central values are given.   The $\dag$ indicates that the particular system does not bind and $?$ denotes the likely existence of a bound state.  }
\label{dk2}
\begin{tabular}{c|ccc|ccc|cccc}
\hline\hline
Molecule  & $I(J^{P})$  & B.E(MeV)  &  Mass(MeV)  & $I(J^{P})$  & B.E(MeV)  &  Mass(MeV) & $I(J^{P})$  & B.E(MeV)  &  Mass(MeV)\\
\hline
${\Sigma}_{c}\Sigma_{c}$  & $0(0^{+})$  &
$39.5^{+19.3}_{-16.2}$ & 4868.5  & $1(1^{+})$     &$1.6^{+2.4}_{-1.3}$   & 4906.4  & $2(0^{+})$   &  $\dag$   & $\dag$ \\
\hline
${\Sigma}_{c}\Sigma_{c}^*-\Sigma_{c}^*\Sigma_{c}$  & $0(1^{+})$  &
$38.4^{+18.6}_{-15.6}$ & $4933.6$  & $1(2^{+})$  &  $2.9^{+3.3}_{-2.1}$ & 4969.1 & $2(1^{+})$ & $\dag$  &$\dag$ \\
${\Sigma}_{c}\Sigma_{c}^*+\Sigma_{c}^*\Sigma_{c}$   & $0(2^{+})$  &
$0.1^{+0.7}_{\dag}$ & $?$ & $1(1^{+})$ & $10.7^{+7.6}_{-5.9}$ & 4961.3 & $2(2^{+})$  & $7.1^{+6.3}_{-4.5}$ & 4964.9 \\
\hline
${\Sigma}_{c}^*\Sigma_{c}^*$  & $0(0^{+})$  &  $59.6^{+25.7}_{-22.2}$& $4976.4$   & $1(1^{+})$  &  $13.7^{+9.0}_{-7.1}$ & 5022.3 & $2(0^{+})$ &  $\dag$ & $\dag$
  \\
${\Sigma}_{c}^*\Sigma_{c}^*$  & $0(2^{+})$  &  $13.1^{+8.4}_{-6.7}$& 5022.9   & $1(3^{+})$  &  $0.4^{+1.2}_{\dag}$ & ? & $2(2^{+})$ &  $0.4^{+1.2}_{\dag}$ & ?
  \\
  \hline\hline
\end{tabular}
\end{table}

 Very recently, Dong et al.  obtained three isoscalar molecules, ${\Sigma}_{c}\Sigma_{c}$, ${\Sigma}_{c}^{\ast}\Sigma_{c}$,  and ${\Sigma}_{c}^{\ast}\Sigma_{c}^{\ast}$ in the single channel Bethe-Salpeter approach,  where they did not take into account the spin-spin interaction~\cite{Dong:2021bvy}. In the same approach,
 Chen et al. have systematically studied the ${\Sigma}_{c}^{(\ast)}\Sigma_{c}^{(\ast)}$ system~\cite{Chen:2021cfl} and found bound states with binding energies larger than ours. In Ref.~\cite{Chen:2021cfl}, the potentials of the ${\Sigma}_{c}^{(\ast)}\Sigma_{c}^{(\ast)}$ system  are determined  by reproducing the masses of $P_{c}(4440)$ and $P_{c}(4457)$, which are assumed as $\bar{D}^{\ast}\Sigma_{c}$ bound states. Since the cutoff of the OBE model in this case should be  1.12 GeV, the binding energies of the ${\Sigma}_{c}^{(\ast)}\Sigma_{c}^{(\ast)}$ system obtained in  Ref~\cite{Chen:2021cfl} are larger than ours.
 A series of theoretical studies of the ${\Sigma}_{c}^{(\ast)}\Sigma_{c}^{(\ast)}$ system~\cite{Lee:2011rka,Li:2012bt,Yang:2018amd,Huang:2013rla,Garcilazo:2020acl,Xia:2021nif,Chen:2021cfl} showed that the predictions on the existence of   ${\Sigma}_{c}^{(\ast)}\Sigma_{c}^{(\ast)}$ molecules are  rather solid, consistent with our results. However, where to search for  the ${\Sigma}_{c}^{(\ast)}\Sigma_{c}^{(\ast)}$ hadronic molecules experimentally has not been thoroughly studied.
 In the following, we investigate the strong decays of likely ${\Sigma}_{c}^{(\ast)}\Sigma_{c}^{(\ast)}$ bound states and hope to stimulate future experimental studies.

\section{Strong decays of ${\Sigma}_{c}^{(\ast)}\Sigma_{c}^{(\ast)}$}
\label{sec:pre}

\begin{figure}[!h]
\begin{center}
\begin{tabular}{ccc}
\begin{minipage}[t]{0.3\linewidth}
\begin{center}
\begin{overpic}[scale=.6]{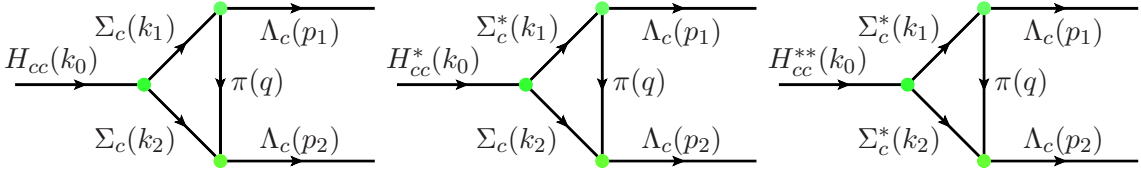}
		\put(68,6){$\Lambda_{c}(p_{2})$}
		
		\put(22,6){$\Sigma_{c}(k_{2})$}
		
		\put(22,37){${\Sigma}_{c}(k_{1})$}
		
		\put(-3,28){$H_{cc}(k_{0})$ }
		\put(68,36){${\Lambda}_{c}(p_{1})$} \put(60,22){$\pi(q)$}
\end{overpic}
\end{center}
\end{minipage}
&
\begin{minipage}[t]{0.3\linewidth}
\begin{center}
\begin{overpic}[scale=.6]{triangle.eps}
		\put(68,6){$\Lambda_{c}(p_{2})$}
		
		\put(22,6){$\Sigma_{c}(k_{2})$}
		
		\put(22,37){${\Sigma}_{c}^{\ast}(k_{1})$}
		
		\put(-3,28){$H_{cc}^{\ast}(k_{0})$ }
		\put(68,36){${\Lambda}_{c}(p_{1})$} \put(60,22){$\pi(q)$}
\end{overpic}
\end{center}
\end{minipage}
&
\begin{minipage}[t]{0.3\linewidth}
\begin{center}
\begin{overpic}[scale=.6]{triangle.eps}
		\put(68,6){$\Lambda_{c}(p_{2})$}
		
		\put(22,6){$\Sigma_{c}^{\ast}(k_{2})$}
		
		\put(22,37){${\Sigma}_{c}^{\ast}(k_{1})$}
		
		\put(-3,28){$H_{cc}^{\ast\ast}(k_{0})$ }
		\put(68,36){${\Lambda}_{c}(p_{1})$} \put(60,22){$\pi(q)$}
\end{overpic}
\end{center}
\end{minipage}
\end{tabular}
\caption{Triangle diagrams of  ${\Sigma}_{c}^{(\ast)}\Sigma_{c}^{(\ast)}$ molecules decaying into ${\Lambda}_{c}\Lambda_{c}$ by exchanging a $\pi$ meson. }
\label{decay34}
\end{center}
\end{figure}

\subsection{Effective Lagrangian approach}

In this section, we further investigate the strong decays of ${\Sigma}_{c}^{(\ast)}\Sigma_{c}^{(\ast)}$ molecules by the effective Lagrangian approach, which has been widely applied to study the strong decays of two-body and three-body bound states~\cite{Faessler:2007gv,Dong:2008gb,Xiao:2019mvs,Dong:2019ofp,Huang:2019qmw,Wu:2020job,Wu:2021gyn}.
The transition from ${\Sigma}_{c}^{(\ast)}\Sigma_{c}^{(\ast)}\to \Lambda_{c}\Lambda_{c}$  is mediated by the exchange of a  $\pi$  meson  {\footnote{In principle, the exchange of  $\rho$ is also allowed. However,   $\Sigma_{c}^{(\ast)}$ can only transit into $\Lambda_{c}\rho$ via the magnetic term, which is regarded as sub-leading for interactions involving vector mesons. As a result,  we can safely neglect the $\rho$ meson exchange in comparison with the  $\pi$ meson exchange.      }}, depicted by the triangle diagrams shown  in Fig.~\ref{decay34}. The  isospin and spin of  the $\Lambda_{c}\Lambda_{c}$ pair are 0 and 0/1, respectively, which can help us select the quantum numbers of initial states because for strong decays isospin and  angular momentum are conserved.  As a result, in this work,  we only study the strong decay of three isoscalar  states,  $J^{P}=0^{+}\Sigma_{c}\Sigma_{c}$, $J^{P}=1^{+}\Sigma_{c}^{\ast}\Sigma_{c}$, and  $J^{P}=0^{+}\Sigma_{c}^{\ast}\Sigma_{c}^{\ast}$,  denoted as $H_{cc}$, $H_{cc}^{\ast}$ and $H_{cc}^{\ast\ast}$ in the following, respectively.

To describe the interaction vertices  of Fig.~\ref{decay34}, we need the following Lagrangians for the ${\Sigma}_{c}^{(\ast)}\Sigma_{c}^{(\ast)}$ bound states and their constituents
\begin{eqnarray}
  \mathcal{L}_{H_{cc}\Sigma_{c}{\Sigma}_{c}}(x)&=&g_{H_{cc}\Sigma_{c}{\Sigma}_{c}}\int dy \Phi(y^2)\bar{H}_{cc}(x)\Sigma_{c}(x+\omega_{\Sigma_{c}} y){\Sigma}_{c}(x-\omega_{\Sigma_{c}} y),
    \\ \nonumber
  \mathcal{L}_{H_{cc}^{\ast}\Sigma_{c}{\Sigma}_{c}^{\ast}}(x)&=&g_{H_{cc}^{\ast}\Sigma_{c}{\Sigma}_{c}^{\ast}}\int dy \Phi(y^2)\bar{H}_{cc}^{\ast}(x)\Sigma_{c}(x+\bar{\omega}_{\Sigma_{c}^{\ast}} y){\Sigma}_{c\nu}^{\ast}(x-\bar{\omega}_{\Sigma_{c}} y),
    \\ \nonumber \mathcal{L}_{H_{cc}^{\ast\ast}\Sigma_{c}^{\ast}\Sigma_{c}^{\ast}}(x)&=&g_{H_{cc}^{\ast\ast}\Sigma_{c}^{\ast}\Sigma_{c}^{\ast}}\int dy \Phi(y^2)\bar{H}_{cc}^{\ast}(x)\Sigma_{c}^{\ast\nu}(x+\omega_{\Sigma_{c}^{\ast}}y){\Sigma}_{c\nu}^{\ast}(x-\omega_{\Sigma_{c}^{\ast}} y),
\end{eqnarray}
where $\omega_{\Sigma_{c}}=\frac{m_{\Sigma_{c}}}{m_{\Sigma_{c}}+m_{\Sigma_{c}}}$, $\omega_{\Sigma_{c}^{\ast}}=\frac{m_{\Sigma_{c}^{\ast}}}{m_{\Sigma_{c}^{\ast}}+m_{\Sigma_{c}^{\ast}}}$, $\bar{\omega}_{\Sigma_{c}^{\ast}}=\frac{m_{\Sigma_{c}^{\ast}}}{m_{\Sigma_{c}}+m_{\Sigma_{c}^{\ast}}}$, and $\bar{\omega}_{\Sigma_{c}}=\frac{m_{\Sigma_{c}}}{m_{\Sigma_{c}^{\ast}}+m_{\Sigma_{c}^{\ast}}}$ are the kinematic parameters with $m_{\Sigma_{c}}$ and $m_{\Sigma_{c}}^{\ast}$ the masses of $\Sigma_{c}$ and $\Sigma_{c}^{\ast}$, and
$g_{H_{cc}\Sigma_{c}{\Sigma}_{c}}$, $g_{H_{cc}^{\ast}\Sigma_{c}{\Sigma}_{c}^{\ast}}$, and $g_{H_{cc}^{\ast\ast}\Sigma_{c}^{\ast}\Sigma_{c}^{\ast}}$ are  the couplings between the ${\Sigma}_{c}^{(\ast)}\Sigma_{c}^{(\ast)}$  bound states  and  their constituents. The $\Phi(y^{2})$ is the correlation function, which not only takes into account the distribution of the two constituent hadrons in a molecule but also renders the Feynman diagrams ultraviolet finite. Here we choose the correlation function in form of a Gaussian function
\begin{eqnarray}
\Phi(y^{2})=\mbox{Exp}(-p_{E}^{2}/\Lambda^2),
\end{eqnarray}
where $p_{E}$ is the Euclidean momentum and $\Lambda$ is the size parameter.

\begin{figure}[ttt]
\centering
\begin{overpic}[scale=.7]{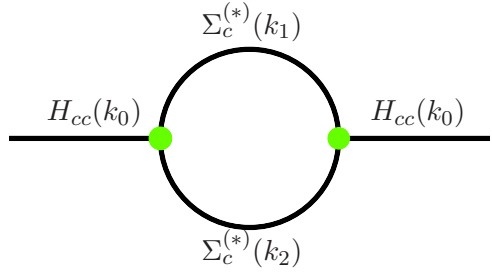}
\put(40,-6){$\Sigma_{c}^{(\ast)}(k_{2})$} \put(40,41){$\Sigma_{c}^{(\ast)}(k_{1})$}
\put(8,23){$H_{cc}(k_{0})$} \put(75,23){$H_{cc}(k_{0})$}
\end{overpic}
\caption{Mass operators of the ${\Sigma}_{c}^{(\ast)}\Sigma_{c}^{(\ast)}$ molecules }
\label{loop}
\end{figure}

To estimate the couplings between bound states and their constituents, we employ the compositeness condition~\cite{Weinberg:1962hj,Salam:1962ap,Hayashi:1967bjx}. For dibaryon bound states with total angular momentum $J=0$ the compositeness condition reads
\begin{eqnarray}
 Z_{\Sigma_{c}\Sigma_{c}}=1-\frac{\partial \Sigma(m_{\Sigma_{c}\Sigma_{c}}^2) }{\partial m_{\Sigma_{c}\Sigma_{c}}^2}=0,\label{composite}
\end{eqnarray}
where $ \Sigma(m_{\Sigma_{c}\Sigma_{c}}^2)$ is the self energy of a ${\Sigma}_{c}^{(\ast)}\Sigma_{c}^{(\ast)}$ bound states as shown in Fig.~\ref{loop}.

The self-energy for a vector state is expressed as $\Sigma^{\mu\nu}(p^2)$ with  Lorentz indices $\mu$ and $\nu$, which can be decomposed into two parts, longitudinal and transverse, $\Sigma^{\mu\nu}(p^2)=\hat{g}^{\mu\nu}\Sigma^{T}(p^{2})+ \frac{p^{\mu}p^{\nu}}{p^{2}}\Sigma^{L}(p^2)$, where $\hat{g}^{\mu\nu}=g^{\mu\nu}-p^{\mu}p^{\nu}/p^2$.  Substituting the transverse term into  Eq~(\ref{composite}), we can also determine the couplings of  a $J=1$ dibaryon state to their constituents.

The mass operators for the ${\Sigma}_{c}^{(\ast)}\Sigma_{c}^{(\ast)}$ dibaryons  read as follows:
\begin{eqnarray}
 M_{H_{cc}}&=&ig_{H_{cc}\Sigma_{c}{\Sigma}_{c}}^{2}\int\frac{d^{4}k_{1}}{(2\pi)^{4}}e^{-\frac{(\omega_{\Sigma_{c}}k_{2}-\omega_{\Sigma_{c}}k_{1})^2}{\Lambda^2}}\frac{1}{{/\!\!\!k_{1}}-m_{\Sigma_c}}\frac{1}{{/\!\!\!k_{2}}-m_{\Sigma_c}},  \\ \nonumber
  M_{H_{cc}^{\ast}}^{\mu\nu}&=&ig_{H_{cc}^{\ast}\Sigma_{c}^{\ast}{\Sigma}_{c}}^{2}\int\frac{d^{4}k_{1}}{(2\pi)^{4}}e^{-\frac{(\bar{\omega}_{\Sigma_{c}^{\ast}}k_{2}-{\bar{\omega}_{\Sigma_{c}}}k_{1})^2}{\Lambda^2}}\frac{P^{\mu\nu}(k_{1})}{{/\!\!\!k_{1}}-m_{\Sigma_c^*}}\frac{1}{{/\!\!\!k_{2}}-m_{\Sigma_c}}, \\ \nonumber
  M_{H_{cc}^{\ast\ast}}&=&ig_{H_{cc}^{\ast\ast}\Sigma_{c}^{\ast}{\Sigma}_{c}^{\ast}}^{2}\int\frac{d^{4}k_{1}}{(2\pi)^{4}}e^{-\frac{(\omega_{\Sigma_{c}^{\ast}}k_{2}-\omega_{\Sigma_{c}^{\ast}}k_{1})^2}{\Lambda^2}}\frac{P^{\lambda\sigma}(k_1)}{{/\!\!\!k_{1}}-m_{\Sigma_c^*}}\frac{P^{\lambda\sigma}(k_{2})}{{/\!\!\!k_{2}}-m_{\Sigma_c^*}},
\end{eqnarray}
where $P^{\lambda\sigma}(p)=g^{\lambda\sigma}-\frac{1}{3}\gamma^{\lambda}\gamma^{\sigma}-\frac{\gamma^{\lambda}p^{\sigma}-\gamma^{\sigma}p^{\lambda}}{3p^2/\slashed{p}}-\frac{2p^{\lambda}p^{\sigma}}{3p^{2}}$.

In Refs.~\cite{Ling:2021lmq,Ling:2021bir}, the above cutoff is  found to be around 1 GeV. One should note, however, that the couplings of a molecular state to its components are related to  its binding energy~\cite{Lin:2019qiv}, and hence we take a cutoff of $\Lambda=0.86$ GeV, the same as that used to study the $\Sigma_{c}^{(*)}\Sigma_c^{(*)}$ dibaryons, to determine the couplings in this work. In Table~\ref{couplings},
we present the  $H_{cc}^{(\ast\ast)}$  couplings to their components. Because the binding energy of $H_{cc}^{\ast\ast}$ is larger than those of $H_{cc}^{\ast}$ and $H_{cc}$, $g_{H_{cc}^{\ast\ast} \Sigma_{c}^{\ast}\Sigma_{c}^{\ast}}$  is larger than $g_{H_{cc}^{\ast} \Sigma_{c}^{\ast}\Sigma_{c}}$ and $g_{H_{cc}  \Sigma_{c}\Sigma_{c}}$  as well.

\begin{table}[ttt]
\centering
\caption{Couplings of    ${\Sigma}_{c}^{(\ast)}\Sigma_{c}^{(\ast)}$ molecules  to their components obtained with a cutoff of $\Lambda=$0.86 GeV.   }
\label{couplings}
\begin{tabular}{c|cccccccccc}
\hline\hline
Couplings  & $g_{H_{cc}\Sigma_{c}{\Sigma}_{c}}$  & $g_{H_{cc}^{\ast}\Sigma_{c}^{\ast}{\Sigma}_{c}}$&  $g_{H_{cc}^{\ast\ast}\Sigma_{c}^{\ast}{\Sigma}_{c}^{\ast}}$ \\
\hline
Value  & 4.3  &
5.2 & 7.7  \\
  \hline\hline
\end{tabular}
\end{table}

The  other vertices of the triangle diagrams of Fig~\ref{decay34}  can be classified into two categories,  $\Sigma_{c}\to \Lambda_{c}\pi$ and $\Sigma_{c}^{\ast}\to \Lambda_{c}\pi$.   The Lagrangians describing these interactions  are given by
\begin{eqnarray}
\mathcal{L}_{\pi\Lambda_{c}{\Sigma}_{c}}&=&-i\frac{g_{\pi\Lambda_{c}{\Sigma}_{c}}}{f_{\pi}}~\bar{\Lambda}_{c}\gamma^\mu\gamma_5 \partial_\mu\vec{\phi}_\pi\cdot \vec{\tau}{\Sigma}_{c},\\ \nonumber
\mathcal{L}_{\pi\Lambda_{c}{\Sigma}_{c}^*}&=&\frac{g_{\pi\Lambda_{c}{\Sigma}_{c}^*}}{f_{\pi}}~\bar{\Lambda}_{c}\partial^\mu\vec{\phi}_\pi\cdot \vec{\tau}{\Sigma}_{c\mu}^*,
\end{eqnarray}
where  $f_{\pi}=132$ MeV and the couplings $g_{\pi\Lambda_{c}{\Sigma}_{c}}$ and $g_{\pi\Lambda_{c}{\Sigma}_{c}^*}$  can be determined by fitting to experimental data. From the decay widths of $\Gamma(\Sigma_{c}\to \Lambda_{c}\pi)=1.89$ MeV and $\Gamma(\Sigma_{c}^{\ast}\to \Lambda_{c}\pi)=15.0$ MeV~\cite{ParticleDataGroup:2018ovx}, we obtain the couplings $g_{\pi\Lambda_{c}{\Sigma}_{c}}=0.55$ and $g_{\pi\Lambda_{c}{\Sigma}_{c}^*}=0.97$,  consistent with other works~\cite{Liu:2011xc,Cheng:2015naa}.
In addition, we find that  the two  couplings approximately satisfy the relationship $g_{\pi\Lambda_{c}{\Sigma}_{c}^*}=\sqrt{3}g_{\pi\Lambda_{c}{\Sigma}_{c}}$, given by the quark model~\cite{Liu:2011xc}.

With the above Lagrangians, the amplitudes of  $H_{cc}^{(\ast\ast)}\to{\Lambda}_{c}\Lambda_{c}$ can be easily written down
\begin{eqnarray}
  \mathcal{M}_{H_{cc}}&=&i\frac{g_{H_{cc}\Sigma_{c}{\Sigma}_{c}}g_{\pi\Lambda_{c}\Sigma_{c}}^{2}}{f_\pi^2}\int\frac{d^{4}q}{(2\pi)^4}\bar{u}_{\Lambda_{c}}\gamma_{\mu}\gamma_{5}q^{\mu}\frac{1}{{/\!\!\!k_{1}}-m_{1}}\frac{1}{q^{2}-m_{\pi}^2}\frac{1}{{/\!\!\!k_{2}}-m_{1}}\gamma_{\nu}\gamma_{5}q^{\nu}F^2(q, m_\pi, \Lambda) \bar{u}_{\Lambda_{c}}^{T},
   \\ \nonumber
  \mathcal{M}_{H_{cc}^{\ast}}&=&i\frac{g_{H_{cc}^{\ast}\Sigma_{c}^{\ast}{\Sigma}_{c}}g_{\pi\Lambda_{c}\Sigma_{c}}g_{\pi\Lambda_{c}\Sigma_{c}^*}}{f_\pi^2}\int\frac{d^{4}q}{(2\pi)^4}\bar{u}_{\Lambda_{c}}q^{\mu}\frac{P^{\mu\nu}(k_{1})\varepsilon_{\nu}(k_{0})}{{/\!\!\!k_{1}}-m_{1}}\frac{1}{q^{2}-m_{\pi}^2}\frac{1}{{/\!\!\!k_{2}}-m_{2}}\gamma_{\alpha}\gamma_{5}q^{\alpha}F^2(q, m_\pi, \Lambda) \bar{u}_{\Lambda_{c}}^{T},
   \\ \nonumber
 \mathcal{M}_{H_{cc}^{\ast\ast}}&=&i\frac{g_{H_{cc}^{\ast\ast}\Sigma_{c}^{\ast}{\Sigma}_{c}^{\ast}}g_{\pi\Lambda_{c}\Sigma_{c}^*}^{2}}{f_\pi^2}\int\frac{d^{4}q}{(2\pi)^4}\bar{u}_{\Lambda_{c}}q^{\mu}\frac{P^{\mu\nu}(k_{1})}{{/\!\!\!k_{1}}-m_{1}}\frac{1}{q^{2}-m_{\pi}^2}\frac{P_{\nu\alpha}(k_2)}{{/\!\!\!k_{2}}-m_{1}}q^{\alpha}F^2(q, m_\pi, \Lambda) \bar{u}_{\Lambda_{c}}^{T},
\end{eqnarray}
where $\bar{u}_{\Lambda_{c}}$ and $\bar{u}_{\Lambda_{c}}^{T}$ represent the spinors of the final-state $\Lambda_{c}\Lambda_c$  pair, and $F(q,m_{E},\Lambda)$ is the form factor
\begin{eqnarray}
F(q,m_{E},\Lambda)=\frac{\Lambda^{2}-m_{E}^2}{\Lambda^2-q^2},
\end{eqnarray}
which not only removes ultraviolet divergence of the loop diagram, but also takes into account the off-shell effects. The $m_{E}$  is the mass of the exchanged particle. The cutoff is expressed by $\Lambda=m_{E}+\alpha \Lambda_{QCD}$, where $\Lambda_{QCD}$ is around 200-300 MeV and the dimensionless parameter $\alpha$ is around 1~\cite{Xiao:2020alj}.  Thus we vary the cutoff from 0.4 GeV to 0.6 GeV to estimate the uncertainties induced.

With the  amplitudes of  $H_{cc}^{(\ast\ast)}\to {\Lambda}_{c}\Lambda_{c}$ determined, one can obtain
 the corresponding partial decay widths  as
 \begin{eqnarray}
\Gamma=\frac{1}{2J+1}\frac{1}{8\pi}\frac{|\vec{p}|}{m_{H_{cc}^{(\ast\ast)}}^2}\bar{|\mathcal{M}|}^{2},
\end{eqnarray}
where $J$ is the total angular momentum of the $H_{cc}^{(\ast\ast)}$ molecule, the overline indicates the sum over the polarization vectors of final states,  $|\vec{p}|$ is the momentum of either final state in the rest frame of  ${\Sigma}_{c}^{(\ast)}\Sigma_{c}^{(\ast)}$ and $m_{H_{cc}^{(\ast\ast)}}$ is the mass of the corresponding molecule. At last, we have to multiply a factor $1/2$ to the above result to take into account the statistics of identical particles.

\subsection{Numerical results and discussions}

\begin{figure}[ttt]
\centering
\begin{overpic}[scale=.45]{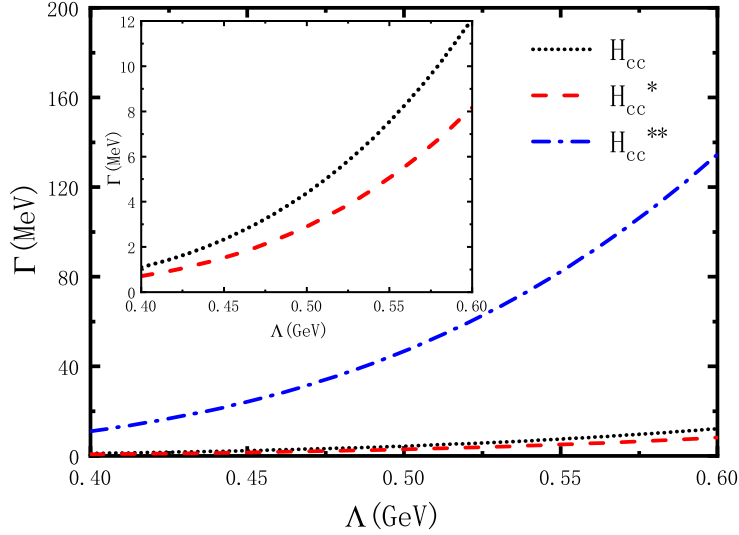}

\end{overpic}
\caption{Decay widths of $H_c^{(**)}\to \Lambda_c^+\Lambda_c^+$ as functions of the cutoff $\Lambda$. }
\label{Gamma}
\end{figure}

In Fig~.\ref{Gamma}, we present the decay widths of $H_{cc}^{(\ast\ast)} \to \Lambda_{c}\Lambda_{c}$ as  functions of the cutoff $\Lambda$, where the masses of  $H_{cc}^{(\ast\ast)}$ molecules are  the central values of Table~\ref{dk2}.  With the cutoff varying from 0.4 to 0.6 GeV, the decay widths of $H_{cc}$ and $H_{cc}^{\ast}$ molecules  change from 1.1 to 12.1 MeV and 0.7 to 8.2 MeV, respectively,  which are close to each other and relatively narrow.  On the other hand, the width of the $H_{cc}^{\ast\ast}$ molecule can be very large, which changes from  11.0 to 134.6 MeV.  For a cutoff of 0.5 GeV, the decay widths of $H_{cc}$, $H_{cc}^{\ast}$, and $H_{cc}^{\ast\ast}$ molecules are 4, 3, and 47 MeV, respectively.

\begin{figure}[!h]
\begin{center}
\begin{tabular}{ccc}
\begin{minipage}[t]{0.34\linewidth}
\begin{center}
\begin{overpic}[scale=.26]{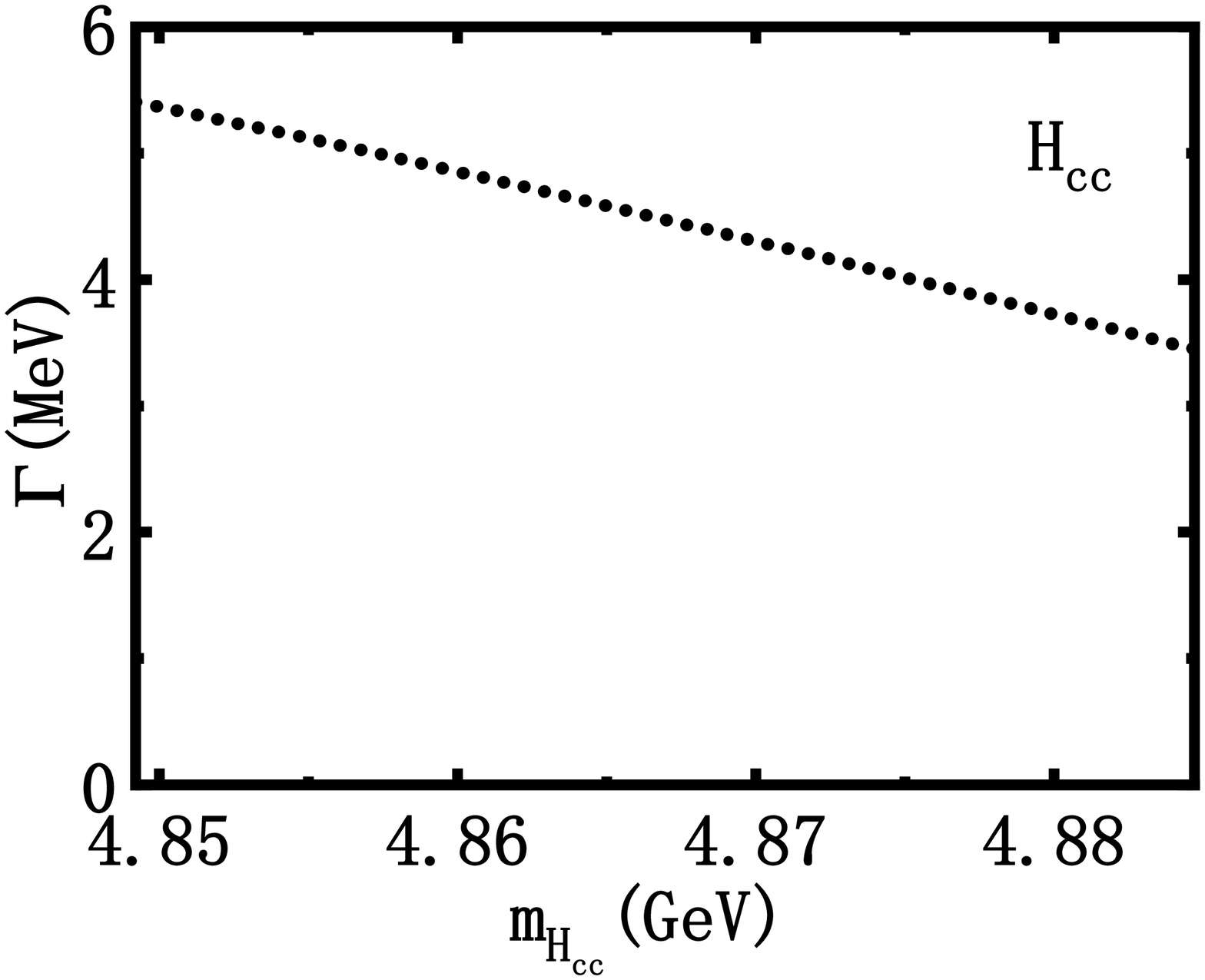}

\end{overpic}
\end{center}
\end{minipage}
&
\begin{minipage}[t]{0.34\linewidth}
\begin{center}
\begin{overpic}[scale=0.26]{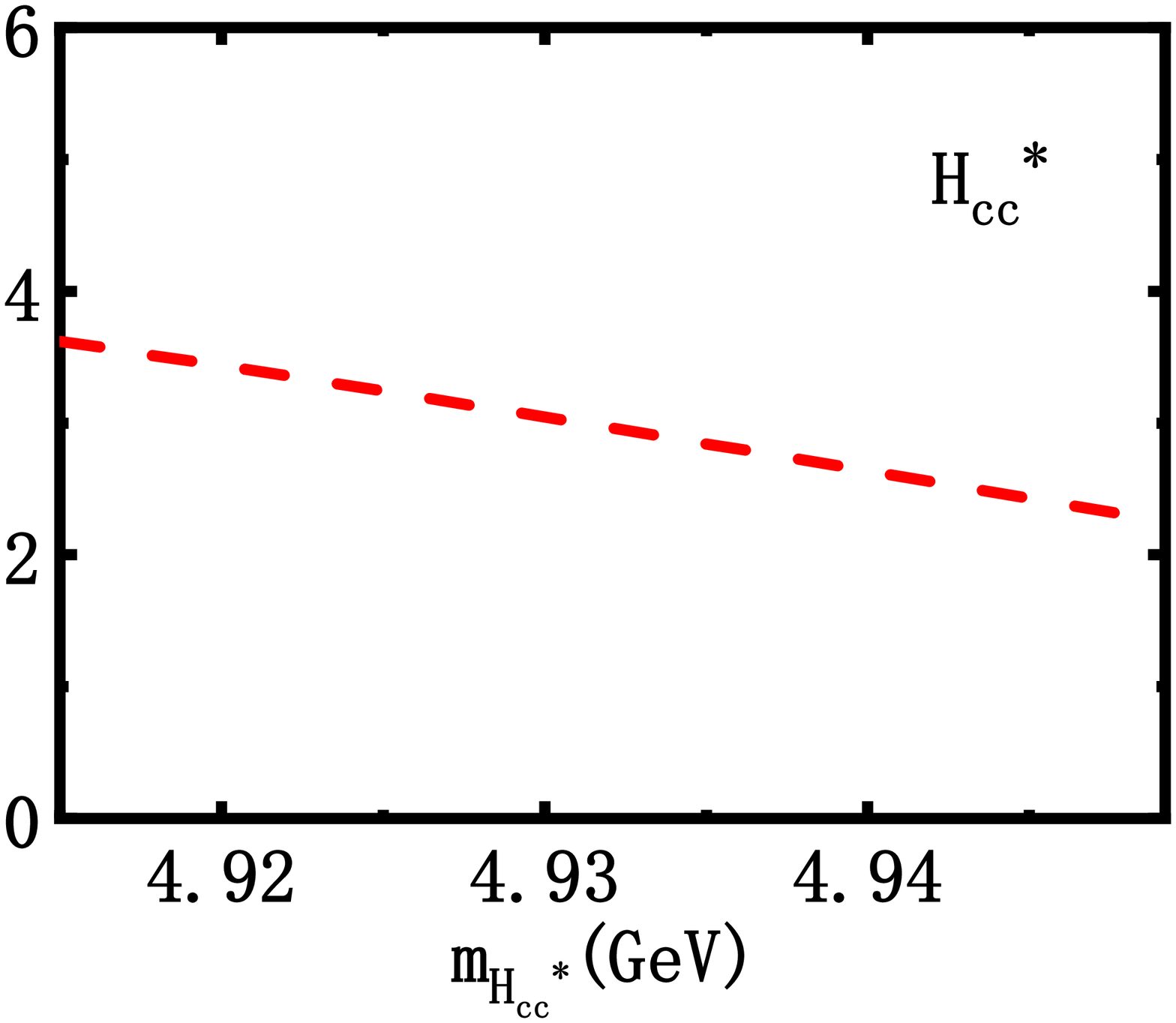}

\end{overpic}
\end{center}
\end{minipage}
&
\begin{minipage}[t]{0.34\linewidth}
\begin{center}
\begin{overpic}[scale=0.26]{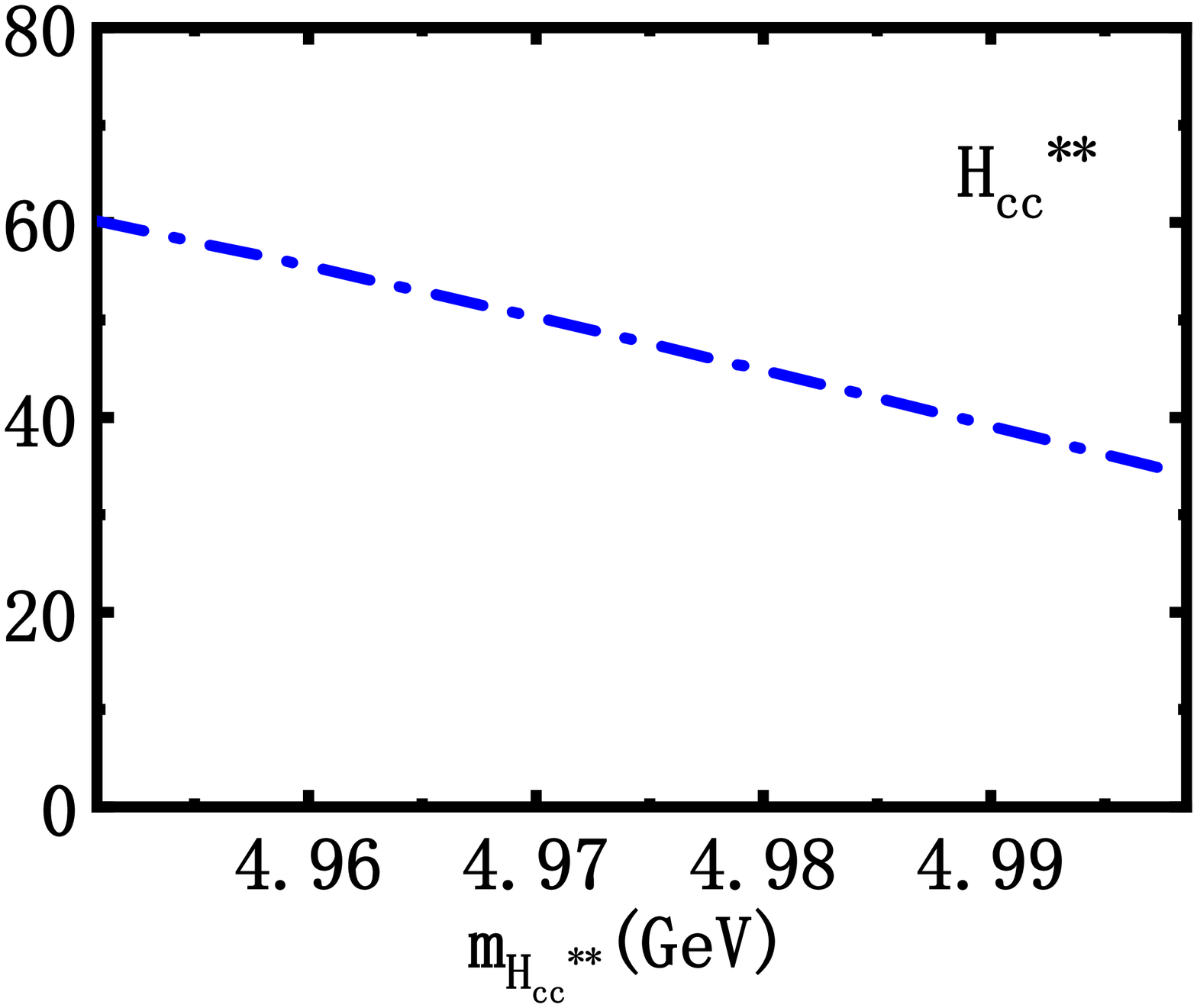}

\end{overpic}
\end{center}
\end{minipage}
\end{tabular}
\caption{Decay widths of $H_c^{(**)}\to \Lambda_c^+\Lambda_c^+$ as functions of the masses of $H_c^{(**)}$, obtained with a cutoff of $\Lambda=0.5$ GeV. }
\label{gamma-m}
\end{center}
\end{figure}

One should note that there also exist uncertainties for the  masses of the  $H_{cc}^{(\ast\ast)}$ molecules in our OBE model. To take into account the impact of mass (binding energy) uncertainties of $H_{cc}^{(\ast\ast)}$ molecules  on the decay widths,
we show the decay widths of  $H_{cc}^{(\ast\ast)}$ molecules as  functions of their masses in Fig.~\ref{gamma-m}, where the ranges of masses are  obtained from the upper  and lower limits  of binding energies of  Table~\ref{dk2}, and the cutoff  is taken to be 0.5 GeV. One can see that the decay widths of $H_{cc}$ and $H_{cc}^{\ast}$ only vary by several MeV, while the decay width of $H_{cc}^{\ast\ast}$ varies by tens of MeV, which shows that the decay width of $H_{cc}^{(\ast\ast)}$ molecules are  not very  sensitive to their masses.

\begin{figure}[!h]
\centering
\begin{overpic}[scale=.45]{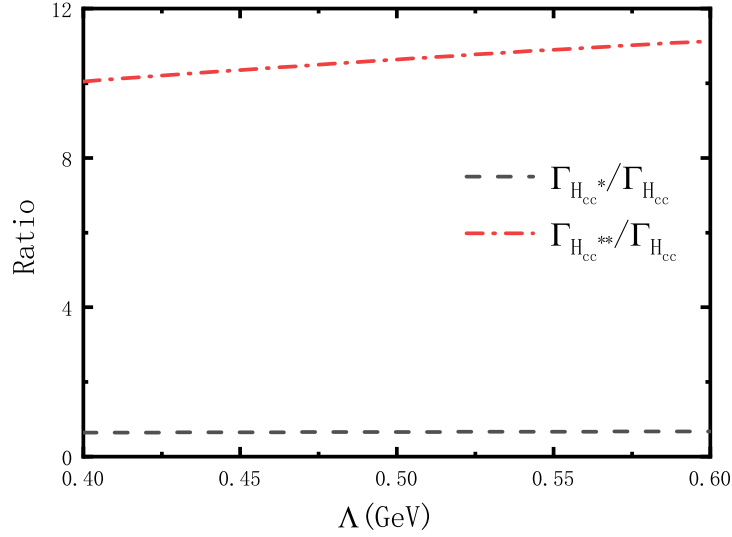}

\end{overpic}
\caption{Ratios between decay widths of $H_c^{(**)}\to \Lambda_c^+\Lambda_c^+$ as functions of the cutoff $\Lambda$. }
\label{Ratio}
\end{figure}

All of the $H_{cc}^{(\ast\ast)} $ molecules can decay into $\Lambda_{c}\Lambda_{c}$, which indicates that all of them could  be detected in the  $\Lambda_{c}\Lambda_{c}$ mass distributions.  In Fig~.\ref{Ratio} we show the ratios of the decay widths of  the $H_{cc}^{\ast\ast}$ and $H_{cc}^{\ast}$ molecules to  that of the $H_{cc}$ molecule, and the corresponding ratios are around 10 and 1, respectively, which tells that  three peaks will appear in the $\Lambda_{c}\Lambda_{c}$ invariant mass spectrum, two narrow structures and one rather wide structure. In addition, we find that these ratios are insensitive to the cutoff used.

\section{Summary and outlook}
\label{sum}

Inspired by the recent discovery of the doubly charmed tetraquark sate $T_{cc}$, we performed a systematic study of the mass spectrum and strong decays of  doubly charmed hexquark states composed of  ${\Sigma}_{c}^{(\ast)}\Sigma_{c}^{(\ast)}$. We adopted the  one-boson exchange model to calculate the binding energies of the ${\Sigma}_{c}^{(\ast)}\Sigma_{c}^{(\ast)}$ system, where the cutoff is fixed by reproducing the binding energy of the deuteron. After considering breaking of HQSS, we found four bound states with isospin 0, i.e., $J^{P}=0^{+}\Sigma_{c}\Sigma_{c}$, $J^{P}=1^{+}\Sigma_{c}^{\ast}\Sigma_{c}$, $J^{P}=0^{+}\Sigma_{c}^{\ast}\Sigma_{c}^{\ast}$, and $J^{P}=2^{+}\Sigma_{c}^{\ast}\Sigma_{c}^{\ast}$, four bound states with isospin 1, $J^{P}=1^{+}$$\Sigma_{c}\Sigma_{c}$, $J^{P}=1^{+}$$\Sigma_{c}^{\ast}\Sigma_{c}$, $J^{P}=2^{+}$$\Sigma_{c}^{\ast}\Sigma_{c}$, and $J^{P}=1^{+}$$\Sigma_{c}^{\ast}\Sigma_{c}^{\ast}$, and one bound state with isospin 2, $J^{P}=2^{+}\Sigma_{c}^{\ast}\Sigma_{c}$. Among them, the $J^{P}=0^{+}\Sigma_{c}\Sigma_{c}$ state could be regarded as the
deuteron counterpart with double charm, which is much more bound than the deuteron. In addition, we found  that with the increase of total isospin, the number of  ${\Sigma}_{c}^{(\ast)}\Sigma_{c}^{(\ast)}$ bound states  decreases  and the binding energies of $I=1$ dibaryons are smaller than those of $I=0$.

All the ${\Sigma}_{c}^{(\ast)}\Sigma_{c}^{(\ast)}$ bound states can decay into $\Lambda_{c}\Lambda_{c}$ by exchange of a $\pi$ meson via the triangle diagrams.    We used the effective Lagrangian approach to estimate the decay widths of ${\Sigma}_{c}^{(\ast)}\Sigma_{c}^{(\ast)}\to \Lambda_{c}\Lambda_{c} $. Due to the conservation law of isospin and spin we only studied  three isoscalar states, i.e., $J^{P}=0^{+}\Sigma_{c}\Sigma_{c}$, $J^{P}=1^{+}\Sigma_{c}^{\ast}\Sigma_{c}$, and  $J^{P}=0^{+}\Sigma_{c}^{\ast}\Sigma_{c}^{\ast}$, which  can decay into  $\Lambda_{c}\Lambda_{c}$. We found that the decay widths of $J^{P}=0^{+}\Sigma_{c}\Sigma_{c}$ and  $J^{P}=1^{+}\Sigma_{c}^{\ast}\Sigma_{c}$ are several MeV and that of   $J^{P}=0^{+}\Sigma_{c}^{\ast}\Sigma_{c}^{\ast}$ is around 
half a hundred MeV,  which depends strongly on the cutoff. We also found that the decay widths of ${\Sigma}_{c}^{(\ast)}\Sigma_{c}^{(\ast)}\to \Lambda_{c}\Lambda_{c} $ are weakly dependent on  the masses of $H_{cc}^{(\ast\ast)}$ molecules.
The ratios of the decay widths of  $H_{cc}^{\ast\ast}$ and $H_{cc}^{\ast}$  to that of $H_{cc}$  are about 10 and 1, respectively, which are only weakly dependent on the cutoff.  We encourage our experimental colleagues to search for the doubly charmed hexaquark states $\Sigma_c^{(*)}\Sigma_c^{(*)}$ in the  $\Lambda_{c}\Lambda_{c}$ invariant mass distributions, which may  be scrutinized at LHC, J-PARC, and RHIC in future.

\section{Acknowledgments}

 This work is supported in part by the National Natural Science Foundation of China under Grants No.11975041,  No.11735003, and No.11961141004. Ming-Zhu Liu acknowledges support from the National Natural Science Foundation of
China under Grants No.1210050997.

\appendix

\begin{table*}[t]
\centering
\centering \caption{Relevant partial wave matrix elements for the $\Sigma_{c}^{(\ast)}\Sigma_{c}^{(\ast)}$ system.} \label{tab:spin1123}
\begin{tabular}{c|c|c|ccccccc}
\hline\hline
State & $J^{P}$  &   Partial wave    & $ \vec{a}_{1}\cdot \vec{a}_{2}$& $3 \vec{a}_1 \cdot \hat{r}\,\vec{a}_2 \cdot \hat{r} -
  \vec{a}_1 \cdot \vec{a}_2$ \\ \hline
 ${\Sigma}_{c}\Sigma_{c}$  &  $J=0$   &   $^1S_{0}$& -3  & 0
\\ \hline
 ${\Sigma}_{c}\Sigma_{c}$  & $J=1$ & $^3S_{1}$-$^3D_{1}$    & $
\left(\begin{matrix}
1 & 0 \\
0 & 1\\
\end{matrix}\right)$
 & $
\left(\begin{matrix}
0 & \sqrt{8} \\
\sqrt{8} & -2\\
\end{matrix}\right)$
\\ \hline
 ${\Sigma}_{c}\Sigma_{c}^{\ast}$   & $J=1$ & $^3S_{1}$-$^3D_{1}$-$^5D_{1}$  &
  $
\left(\begin{matrix}
-\frac{5}{2} & 0 & 0\\
0 & -\frac{5}{2}& 0\\
  0    &0          &\frac{3}{2}  \\
\end{matrix}\right)$  &  $
\left(\begin{matrix}
0 & -\frac{1}{\sqrt{2}} & -\frac{3}{\sqrt{2}}\\
-\frac{1}{\sqrt{2}}  & \frac{1}{2}& -\frac{3}{2}\\
 -\frac{3}{\sqrt{2}}    &-\frac{3}{2}        &-\frac{3}{2}  \\
\end{matrix}\right)$
\\ \hline
 ${\Sigma}_{c}\Sigma_{c}^{\ast}$   & $J=2$ & $^5S_{2}$-$^3D_{2}$-$^5D_{2}$  & $
\left(\begin{matrix}
\frac{3}{2} & 0 & 0\\
0 & -\frac{5}{2}& 0\\
  0    &0          &\frac{3}{2}  \\
\end{matrix}\right)$  &  $
\left(\begin{matrix}
0 & 3\sqrt{\frac{3}{10}} & 3\sqrt{\frac{7}{10}}\\
3\sqrt{\frac{3}{10}}   & -\frac{1}{2}& -\frac{3}{2}\sqrt{\frac{3}{7}}\\
3\sqrt{\frac{7}{10}}    &-\frac{3}{2}\sqrt{\frac{3}{7}}      &\frac{9}{14}  \\
\end{matrix}\right)$
\\ \hline
 ${\Sigma}_{c}^{\ast}\Sigma_{c}^{\ast}$  & $J=0$ & $^1S_{0}$-$^5D_{0}$
& $
\left(\begin{matrix}
-\frac{15}{4} & 0 \\
0 & -\frac{3}{4}\\
\end{matrix}\right)$  &$
\left(\begin{matrix}
0 & -3 \\
-3 & -3\\
\end{matrix}\right)$   \\ \hline
 ${\Sigma}_{c}^{\ast}\Sigma_{c}^{\ast}$  & $J=1$ & $^3S_{1}$-$^3D_{1}$-$^5D_{1}$-$^7D_{1}$  &  $
\left(\begin{matrix}
-\frac{11}{4} & 0  &0 &0\\
0 & -\frac{11}{4} & 0  &0\\
0 & 0&  -\frac{3}{4} & 0  \\
0 & 0&  0&  \frac{9}{4} \\
\end{matrix}\right)$  &$
\left(\begin{matrix}
0 & \frac{17}{5\sqrt{2}}  &0 &-\frac{3\sqrt{7}}{5}\\
\frac{17}{5\sqrt{2}} & -\frac{17}{10} & 0  &\frac{3}{5}\sqrt{\frac{2}{7}}\\
0 & 0&  -\frac{3}{2} & 0  \\
-\frac{3\sqrt{7}}{5} & \frac{3}{5}\sqrt{\frac{2}{7}}&  0 &  -\frac{108}{35} \\
\end{matrix}\right)$   \\ \hline
 ${\Sigma}_{c}^{\ast}\Sigma_{c}^{\ast}$  & $J=2$ & $^5S_{2}$-$^1D_{2}$-$^3D_{2}$-$^5D_{2}$-$^7D_{2}$   & $
\left(\begin{matrix}
-\frac{3}{4} & 0  &0 &0 &0\\
0 & -\frac{15}{4} & 0  &0  &0\\
0 & 0&  -\frac{11}{4} & 0 &0 \\
0 & 0&  0&  -\frac{3}{4} &0\\
0 & 0&  0&  0& \frac{9}{4} \\
\end{matrix}\right)$  &$
\left(\begin{matrix}
0 & -\frac{3}{\sqrt{5}}  &0 &3\sqrt{\frac{7}{10}} &0\\
-\frac{3}{\sqrt{5}}  & 0 & 0  & 3\sqrt{\frac{2}{7}}  &0\\
0 & 0&  \frac{17}{10} & 0 &\frac{6}{5}\sqrt{\frac{3}{7}} \\
3\sqrt{\frac{7}{10}} & 3\sqrt{\frac{2}{7}}&  0&  \frac{9}{14} &0\\
0 & 0&   \frac{6}{5}\sqrt{\frac{3}{7}}& 0 & -\frac{27}{35} \\
\end{matrix}\right)$
\\ \hline
 ${\Sigma}_{c}^{\ast}\Sigma_{c}^{\ast}$  & $J=3$ & $^7S_{3}$-$^3D_{3}$-$^5D_{3}$-$^7D_{3}$  & $
\left(\begin{matrix}
\frac{9}{4} & 0  &0 &0\\
0 & -\frac{11}{4} & 0  &0\\
0 & 0&  -\frac{3}{4} & 0  \\
0 & 0&  0&  \frac{9}{4} \\
\end{matrix}\right)$  &$
\left(\begin{matrix}
0 & -\frac{3\sqrt{3}}{5}  &0 &\frac{5\sqrt{3}}{9}\\
-\frac{3\sqrt{3}}{5}  & -\frac{17}{35} & 0  &\frac{16}{35}\\
0 & 0&  \frac{12}{7} & 0  \\
\frac{5\sqrt{3}}{9} & \frac{16}{35}&  0 &  \frac{99}{70} \\
\end{matrix}\right)$
\\ \hline\hline
\end{tabular}
\end{table*}

\bibliography{XiccSigmac}

\end{document}